\documentclass[oldversion]{aa}
\usepackage{amsmath}
\usepackage{txfonts}
\usepackage{epsfig,graphicx}
\usepackage{natbib}
\bibpunct{(}{)}{;}{a}{}{,}

\begin{document}
   \title{Spectroscopic parameters for 451 stars in the HARPS GTO planet search program\thanks{Based on observations collected
       at La Silla Observatory, ESO, Chile, with the HARPS spectrograph
       at the 3.6-m telescope (072.C-0488(E))}}

   \subtitle{Stellar [Fe/H] and the frequency of exo-Neptunes}

   \author{S. G. Sousa\inst{1,}\inst{2}, N. C. Santos\inst{1,}\inst{3}, M. Mayor\inst{3}, S. Udry\inst{3}, L. Casagrande\inst{4}, G. Israelian \inst{5} , F. Pepe\inst{3}, D. Queloz\inst{3},  M. J. P. F. G. Monteiro\inst{1,}\inst{2}    }

   \offprints{S. G. Sousa: sousasag@astro.up.pt}

   \institute{Centro de Astrof\'isica, Universidade do Porto, Rua das Estrelas, 4150-762 Porto, Portugal
\and Departamento de Matem\'atica Aplicada, Faculdade de Ci\^encias da Universidade do Porto, Portugal
\and Observatoire de Gen\`eve, 51 Ch. des Mailletes, 1290 Sauverny, Switzerland
\and University of Turku - Tuorla Astronomical Observatory, V\"ais\"al\"antie 20, FI-21500 Piikki\"o, Finland
\and Instituto de Astrof\'isica de Canarias, 38200 La Laguna, Tenerife, Spain
     }

   \date{Received <date>; accepted <date>}


  \abstract{
   {}
   {To understand the formation and evolution of solar-type stars in the solar neighborhood, we need to measure their stellar parameters to high accuracy.
   }
   {We present a catalogue of accurate stellar parameters for 451 stars that represent the HARPS Guaranteed Time Observations (GTO) ``high precision'' sample. Spectroscopic stellar parameters were measured using high signal-to-noise (S/N) spectra acquired with the HARPS spectrograph. The spectroscopic analysis was completed assuming LTE with a grid of Kurucz atmosphere models and the recent ARES code for measuring line equivalent widths.}
   {We show that our results agree well with those ones presented in the literature (for stars in common). We present a useful calibration for the effective temperature as a function of the index color \textit{B-V} and [Fe/H]. We use our results to study the metallicity-planet correlation, namely for very low mass planets.}
   {The results presented here suggest that in contrast to their jovian couterparts, neptune-like planets do not form preferentially
   around metal-rich stars. The ratio of jupiter-to-neptunes is also an increasing function of stellar metallicity.
   These results are discussed in the context of the core-accretion model for planet formation.
   }}


\authorrunning{Sousa, S. G. et al.}

\titlerunning{Spectroscopic parameters for 451 stars in the HARPS GTO planet search program}

\maketitle


\section{Introduction}

Extra-solar planets have continued to be discovered, since the first detection of a giant planet orbiting a solar-like star \citep[][]{Mayor-1995}, including several planets in the Neptune-mass regime \citep[e.g.][]{McArthur-2004, Santos-2004b, butler-2004, Vogt-2005, Rivera-2005, Bonfils-2005, Bonfils-2007, Udry-2006, Udry-2007, Lovis-2006, Melo-2007}. This has been possible because of to the development of a new generation of instruments capable of radial-velocity measurements of unprecedented quality. One such instrument is undoubtedly the ESO high-resolution HARPS fiber-fed echelle spectrograph, especially designed for planet-search programmes and asteroseismology. HARPS has proved to be the most precise spectro-velocimeter to date, which has achieved an instrumental radial-velocity accuracy superior to 1 $m s^{-1}$ for many years \citep[][]{Mayor-2003, Lovis-2005}.


A by-product of these planet searches are the high-quality spectra, of high S/N and high-resolution, that are collected in every run in a consistent and systematic way for selected candidate stars. The result is a vast quantity of spectra, mainly of solar-type stars, which can be analyzed in a consistent way for the determination of accurate spectroscopic stellar parameters.

We present a catalogue of spectroscopic stellar parameters for the HARPS Guaranteed Time Observations (GTO) ``high precision'' sample. In Sect. 2 we describe the observations with HARPS. Section 3 describes the procedure used to derive accurate spectroscopic stellar parameters. In Sect. 4, we present the results for our sample of 451 solar-type stars. We then compare, in Sect. 5, our results with those of several other authors obtained using different methods. In Sect. 6, we present a useful calibration of the effective temperature as a function of $B-V$ and [Fe/H]. Finally, we perform a simple statistical analysis of our results in the context of the metallicity-planet correlation.

\section{Stellar Sample \& Observations}

The objective of the HARPS ``high-precision'' GTO project is to detect very low-mass extra-solar planets by increasing the radial velocity measurement accuracy below 1 $m s^{-1}$. The stars in this project were selected from a volume-limited stellar-sample observed by the CORALIE spectrograph at La Silla observatory \citep[][]{Udry-2000}. These stars are slowly-rotating, non-evolved, and low-activity stars that presented no obvious radial-velocity variations at the level of the CORALIE measurement precision (typically 5-10 $m s^{-1}$ and dominated by photon-noise). Southern planet hosts were also added to this program to search for further low-mass planets. Therefore, a significant number of stars (66) from this catalogue are already confirmed planet hosts. For more details we point the reader to a description of this sample by \citet[][]{Mayor-2003}. In summary, the sample is composed of 451 FGK stars with apparent magnitudes that range from 3.5 to 10.2 and have distances less than 56 parsec.

Asteroseismology work using HARPS has proved that the precision of the instrument is no longer set by instrumental characteristics but rather by the stars themselves \citep[][]{Mayor-2003, Bouchy-2005}. Even a ``quiet'' G dwarf shows oscillation modes of several tens of $cm s^{-1}$ each, which might add up to radial-velocity amplitudes as large as several $m s^{-1}$ \citep[][]{Bouchy-2005}. As a consequence, any exposure with a shorter integration time than the oscillation period of the star, might fall arbitrarily on a peak or on a valley of these mode interferences and thus introduce additional radial-velocity ``noise''. This phenomenon could seriously compromise the ability to detect very low-mass planets around solar-type stars by means of the radial-velocity technique.

To average out stellar oscillations, the observations are designed to last at least 15 minutes on the target, splitting them into several exposures as required, to avoid CCD saturation. The final result is a high-quality spectrum for each star in the sample. 

The individual spectra of each star were reduced using the HARPS pipeline and combined using IRAF \footnote{IRAF is distributed by National Optical Astronomy Observatories, operated by the Association of Universities for Research in Astronomy, Inc., under contract with the National Science Foundation, U.S.A.} after correcting for its radial velocity. The final spectra have a resolution of R $\sim$ 110\,000 and signal-to-noise ratios that vary from $\sim 70$ to $\sim 2000$, depending on the amount and quality of the original spectra (90\% of the spectra have S/N higher than 200).

\section{Stellar parameters - Spectroscopic Analysis}

We followed the procedure presented in \citet[][]{Santos-2004b} for deriving stellar parameters and metallicities, where [Fe/H] is used as a proxy for overall stellar metallicity. As reported in \citet[][]{Sousa-2007} however, we attempted to increase the number of spectral lines and decrease the errors involved in the determination of stellar parameters. Therefore, we used the large, extended line-list presented in \citet[][]{Sousa-2007} and applied this method to the sample.

The spectroscopic analysis was completed assuming LTE, and using the 2002 version of the code MOOG\footnote{The source code of MOOG2002 can be
downloaded at http://verdi.as.utexas.edu/moog.html} \citep[][]{Sneden-1973} and a grid of Kurucz Atlas\,9 plane-parallel model atmospheres \citep[][]{Kurucz-1993}. Stellar parameters and metallicities were derived as in previous works \citep[][]{Santos-2004b,Santos-2005,Sousa-2006}, based on the equivalent widths of \ion{Fe}{i} and \ion{Fe}{ii} weak lines, by imposing excitation and ionization equilibrium.

\begin{table}[t]
\caption[]{Examples of atomic parameters and measured solar equivalent widths for
\ion{Fe}{ii} and \ion{Fe}{i} lines.}
\begin{tabular}{lcccc}
\hline
\hline
\noalign{\smallskip}
$\lambda$ (\AA) & $\chi_{l}$ & $\log{gf}$ & Ele. & EW$_{\sun}$ (m\AA)\\
\hline
4508.28  &   2.86  &   -2.403 & FeII  &   87.3 \\
4520.22  &   2.81  &   -2.563 & FeII  &   81.9 \\
4523.40  &   3.65  &   -1.871 & FeI   &   44.2 \\
4531.62  &   3.21  &   -1.801 & FeI   &   66.4 \\
4534.17  &   2.86  &   -3.203 & FeII  &   53.7 \\
4537.67  &   3.27  &   -2.870 & FeI   &   17.4 \\
4541.52  &   2.86  &   -2.762 & FeII  &   71.5 \\
4551.65  &   3.94  &   -1.928 & FeI   &   29.1 \\
4554.46  &   2.87  &   -2.752 & FeI   &   37.4 \\
4556.93  &   3.25  &   -2.644 & FeI   &   26.3 \\
4576.34  &   2.84  &   -2.947 & FeII  &   64.9 \\
  ...    &    ...  &     ...  & ...   &   ...  \\
\hline
\end{tabular}
\label{tab1}
\end{table}

\subsection{A large list of stable lines}

We considered the line list used for the spectroscopic analysis given in \citet[][]{Sousa-2007}. As a first step, to check the quality of the line list, we derived preliminary stellar parameters for our sample. The stability of the lines were then checked by comparing the abundances of each individual line with the mean abundance obtained for each star.

\begin{figure}[htp]
\centering
\includegraphics[width=8cm]{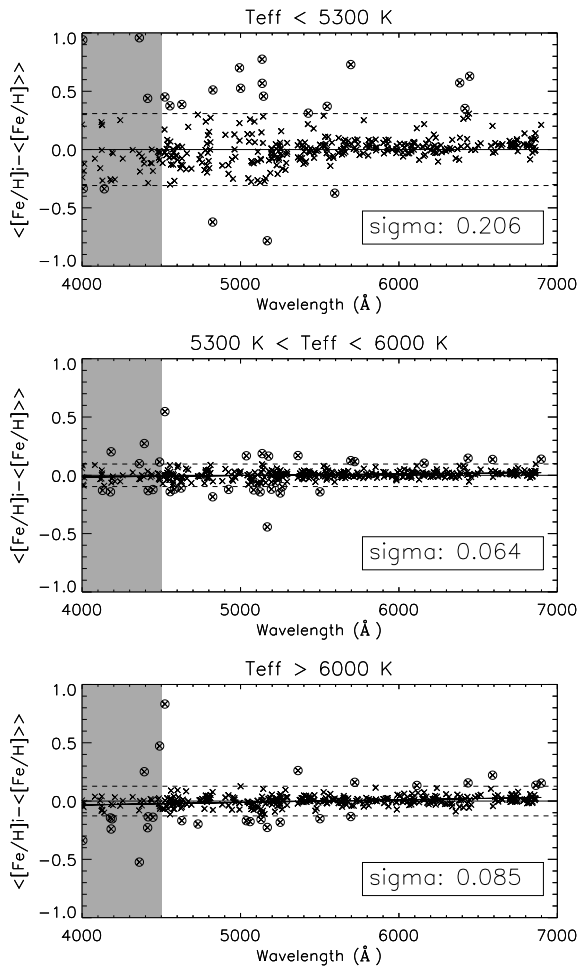}
\caption[]{Selection of the stable lines from the list provided by \citet[][]{Sousa-2007}. We present the mean difference between each line abundance and the mean abundance of the respective star as a function of the wavelength. We divided the analysis into three different ranges of effective temperatures. The dashed lines represent the 1.5 sigma threshold. Lines in the shaded area ($< 4500$\AA) were also removed since this region was strongly affected by blending effects, in particular for cooler stars.}
\label{fig_linelist}
\end{figure}

Since the stars in our sample have a significant range of spectral type with effective temperatures ranging from $\sim$4500 K to $\sim$6400 K, we expect a significant different local environment around each line, due to variations in blending effects. To take this problem into account, we divided our sample into three subsamples, taking into consideration the preliminary determination of the effective temperature ($T_{\mathrm{eff}} < 5300K$, $5300K < T_{\mathrm{eff}} < 6000 K$, $T_{\mathrm{eff}} > 6000K$).

This allows the search for systematic non-stable lines in the large list for the three ranges of stellar spectral class. The lines that showed a systematic offset from the mean value, that was larger than 1.5 sigma were considered ``bad'' lines, which provide false abundance measurements either due to line-blending events or because the continuum could not be fitted readily in a particular spectral region. Lines at wavelengths lower than 4500\AA\ were removed because, for cooler stars, this region is highly line crowded and therefore more likely to produce blending effects. These lines were removed from the original large line list.

Table \ref{tab1} shows the final list composed of 263 \ion{Fe}{i} and 36 \ion{Fe}{ii} weak lines to be used on HARPS spectra for the determination of the stellar parameters and iron abundance.

We note that this procedure is difficult to apply to cooler stars, because the spectra of these stars are crowded with lines or the atomic parameters that we adopt become inaccurate as we study spectral classes increasingly different from solar. These two effects can be seen clearly in the dispersion of measurements shown in Fig. \ref{fig_linelist} (upper panel) that is significantly higher than for the two other cases (middle and lower panel).

Due to these problems, the solutions for cooler stars are more difficult to converge into the ``true'' stellar parameters. In a few cases, we had to discard some lines from the final stellar-line list with equivalent widths greater that 200 m\AA\ and smaller than 10m\AA\ to overcome this issue.

\subsection{Using ARES}

The equivalent widths of the lines were measured automatically in all spectra using the ARES\footnote{The ARES code can be downloaded at http://www.astro.up.pt/$\sim$sousasag/ares} code \citep[Automatic Routine for line Equivalent widths in stellar Spectra - ][]{Sousa-2007}. 

The ARES code considers a one-dimensional spectrum, a list of absorption lines to be measured, and some calibration parameters. First, ARES estimates the continuum about each individual line and calculates a local normalization. The location and number of lines for which a Gaussian fit can be attempted, are then determined. After fitting the local, normalized spectrum, ARES uses the fit parameters to compute the equivalent width. The result is then outputted to a file.

A number of ARES parameters were fixed during the analysis of all spectra in our sample: $smoothder = 4$, $space = 3$, $lineresol = 0.07$, $miniline = 2$. The \textit{smoothder} value was used to reduce the noise in the computed derivatives of the local spectrum to find the number and location of the lines to be fitted. The \textit{space} parameter was set to be the total length of spectrum about each line required to fit the local continuum; therefore, for a value of 3 \AA, ARES fitted the continuum over a total interval of 6 \AA. The \textit{lineresol} parameter defines the minimum separation allowed for consecutive lines. This parameter is used to limit the effect of noise on the determination of the number and location of lines, i.e. if ARES identify lines that are close together according to the value indicated in \textit{lineresol}, then it ignores one line and then measures a mean position. The \textit{miniline} parameter is the lower value of EW that will be returned by ARES in the output file.

\begin{center}
\begin{table}[th]
\caption[]{Dependence of the \textit{rejt} parameter on the S/N of HARPS spectra. ARES users should use this table as a reference for the rejt parameter.}
\begin{tabular}{cc|cc}
\hline
\hline
\noalign{\smallskip}
S/N condition   & \textit{rejt}&            S/N condition        & \textit{rejt}\\
\hline
~~~~~~~~~~   S/N $<$ 100 	&	0.985 & 				&	       \\
100~  $\le$ S/N $<$ 125 	&	0.990 & 250~  $\le$ S/N $<$ 300~	&	0.995  \\
125~  $\le$ S/N $<$ 150 	&	0.991 & 300~  $\le$ S/N $<$ 500~	&	0.996  \\
150~  $\le$ S/N $<$ 200 	&	0.992 & 500~  $\le$ S/N $<$ 800~	&	0.997  \\
200~  $\le$ S/N $<$ 225 	&	0.993 & 800~  $\le$ S/N $<$ 1500	&	0.998  \\
225~  $\le$ S/N $<$ 250 	&	0.994 & 1500  $\le$ S/N ~~~~~~~~~~~~    &	0.999  \\
\hline
\end{tabular}
\label{tab2}
\end{table}
\end{center}

\begin{table*}[t]
\caption[]{Sample table of the HARPS GTO ``high precision'' spectroscopic catalogue. We provide the star name, effective temperature, surface gravity, micro turbulence, [Fe/H], the number of iron lines used in the spectroscopic analysis, a mass estimate, the surface gravity computed from the parallaxes, the luminosity of the star, and an indication of whether it hosts a planet. In the electronic table, we also indicate the error in the luminosity and the computed absolute magnitude of the stars.}
\begin{tabular}{lcccrccccc}
\hline
\hline
Star ID     & T$_{\mathrm{eff}}$ & $\log{g}_{spec}$ & $\xi_{\mathrm{t}}$ & \multicolumn{1}{c}{[Fe/H]} & N(\ion{Fe}{i},\ion{Fe}{ii}) & Mass       & $\log{g}_{hipp}$ & Luminosity & planet host?\\
            & [K]                & [cm\,s$^{-2}$]   &  [km\,s$^{-1}$]    &                            &                             & [M$_{\sun}$]     & [cm\,s$^{-2}$] &   [$L_{\sun}$]   &  \\
\hline

...                    &        ...  &  ...  &  ...  &   ...  &  ...  &  ...  &   ...    & ...     &...\\
\object{HD\,63454} &            4840   \ $\pm$\      66  & 4.30   \ $\pm$\     0.16 &  0.81  \ $\pm$\     0.14 &   0.06\ $\pm$\ 0.03  &  227, 27  &   0.66     &        4.55      &     0.25    &      yes   \\
\object{HD\,63765} &             5432   \ $\pm$\      19  & 4.42   \ $\pm$\     0.03 &  0.82  \ $\pm$\     0.03 &  -0.16\ $\pm$\ 0.01  &  249, 32  &   0.83     &        4.51     &      0.55    &       no    \\
\object{HD\,65216 } &            5612  \ $\pm$\       16  & 4.44    \ $\pm$\    0.02 &  0.78  \ $\pm$\     0.03 &  -0.17\ $\pm$\ 0.01  &  256, 34  &   0.87     &        4.48     &      0.71    &      yes   \\
\object{HD\,65277} &             4802   \ $\pm$\      88  & 4.43    \ $\pm$\    0.18 &  0.55  \ $\pm$\     0.34 &  -0.31\ $\pm$\ 0.04  &  257, 32   &  0.49     &        4.49     &      0.21     &      no    \\
\object{HD\,65562} &             5076   \ $\pm$\      47  & 4.39    \ $\pm$\    0.09 &  0.45  \ $\pm$\     0.18 &  -0.10\ $\pm$\ 0.03  &  252, 34   &  0.72     &        4.51      &     0.37     &      no    \\
\object{HD\,65907A} &            5945   \ $\pm$\      16  & 4.52   \ $\pm$\     0.02 &  1.05  \ $\pm$\     0.02 &  -0.31\ $\pm$\ 0.01 &   256, 34  &   0.93     &        4.37      &     1.24    &       no    \\
\object{HD\,66221} &             5635   \ $\pm$\      25  & 4.40    \ $\pm$\    0.04 &  0.92  \ $\pm$\     0.03  &  0.17\ $\pm$\ 0.02  &  260, 35  &   0.96     &        4.40      &     0.95     &      no   \\ 
\object{HD\,66428} &             5705   \ $\pm$\      27  & 4.31    \ $\pm$\    0.06 &  0.96  \ $\pm$\     0.03 &   0.25\ $\pm$\ 0.02  &  258, 35  &   1.01      &       4.31      &     1.28     &     yes   \\
\object{HD\,67458} &             5891   \ $\pm$\      12  & 4.53    \ $\pm$\    0.02 &  1.04  \ $\pm$\     0.01 &  -0.16\ $\pm$\ 0.01  &  256, 35   &  0.98      &       4.45     &      1.02    &       no    \\
\object{HD\,68607 } &            5215    \ $\pm$\     45  & 4.41    \ $\pm$\    0.08 &  0.82  \ $\pm$\     0.08  &  0.07\ $\pm$\ 0.03  &  251, 35  &   0.81      &       4.52     &      0.45    &       no    \\
\object{HD\,68978A} &            5965   \ $\pm$\      22  & 4.48    \ $\pm$\    0.02  & 1.09  \ $\pm$\     0.02  &  0.05\ $\pm$\ 0.02  &  260, 34   &  1.08      &       4.43     &      1.24    &       no    \\
\object{HD\,69655 } &            5961   \ $\pm$\      12  & 4.44     \ $\pm$\   0.03  & 1.15  \ $\pm$\     0.01 &  -0.19\ $\pm$\ 0.01  &  249, 34  &   0.98      &       4.36     &      1.33    &       no    \\
\object{HD\,69830} &             5402   \ $\pm$\      28  & 4.40     \ $\pm$\   0.04  & 0.80  \ $\pm$\     0.04 &  -0.06\ $\pm$\ 0.02  &  255, 36  &   0.82       &      4.46     &      0.60    &      yes   \\
\object{HD\,70642 } &            5668   \ $\pm$\      22  & 4.40     \ $\pm$\   0.04  & 0.82  \ $\pm$\     0.03  &  0.18\ $\pm$\ 0.02  &  253, 36  &   0.98       &      4.42     &      0.95    &      yes   \\
...                    &        ...  &  ...  &  ...  &   ...  &  ...  &  ...  &   ...    & ...     &...\\

\hline
\end{tabular}

\label{tab3}
\end{table*}

As described in \citet[][]{Sousa-2007}, the most important parameter for the correct automatic determination of equivalent widths using ARES is the \textit{rejt} parameter. This parameter strongly depends on the signal-to-noise ratio (S/N) of the spectra. Values close to 1 are indicative of lower spectrum noise levels. Table \ref{tab2} shows how the parameter was defined by considering the spectrum S/N (measured at 6050 \AA). Future ARES users may use this table as a reference for the most appropriate choice of the \textit{rejt} parameter. This table was compiled for HARPS spectra, but the data should be applicable to other high-resolution spectra.

\subsection{Stellar mass and luminosity}

Stellar masses were estimated, as in previous works e.g. \citet[][]{Santos-2004b}, by interpolating the theoretical isochrones of \citet[][]{Schaller-1992} and \citet[][]{Schaerer-1993b,Schaerer-1993a}, using $M_V$ computed for the Hipparcos parallaxes and V magnitudes \citep[][]{ESA-1997}, a bolometric correction from \citet[][]{Flower-1996}, and the effective temperature derived from the spectroscopic analysis. We adopt a typical relative error of 0.10 $M_{\sun}$ for the masses. In some cases, the value of mass was not determined because the calculation would have required a large extrapolation of the existing isochrones.

\begin{figure}[h]
\centering
\includegraphics[width=6.9cm]{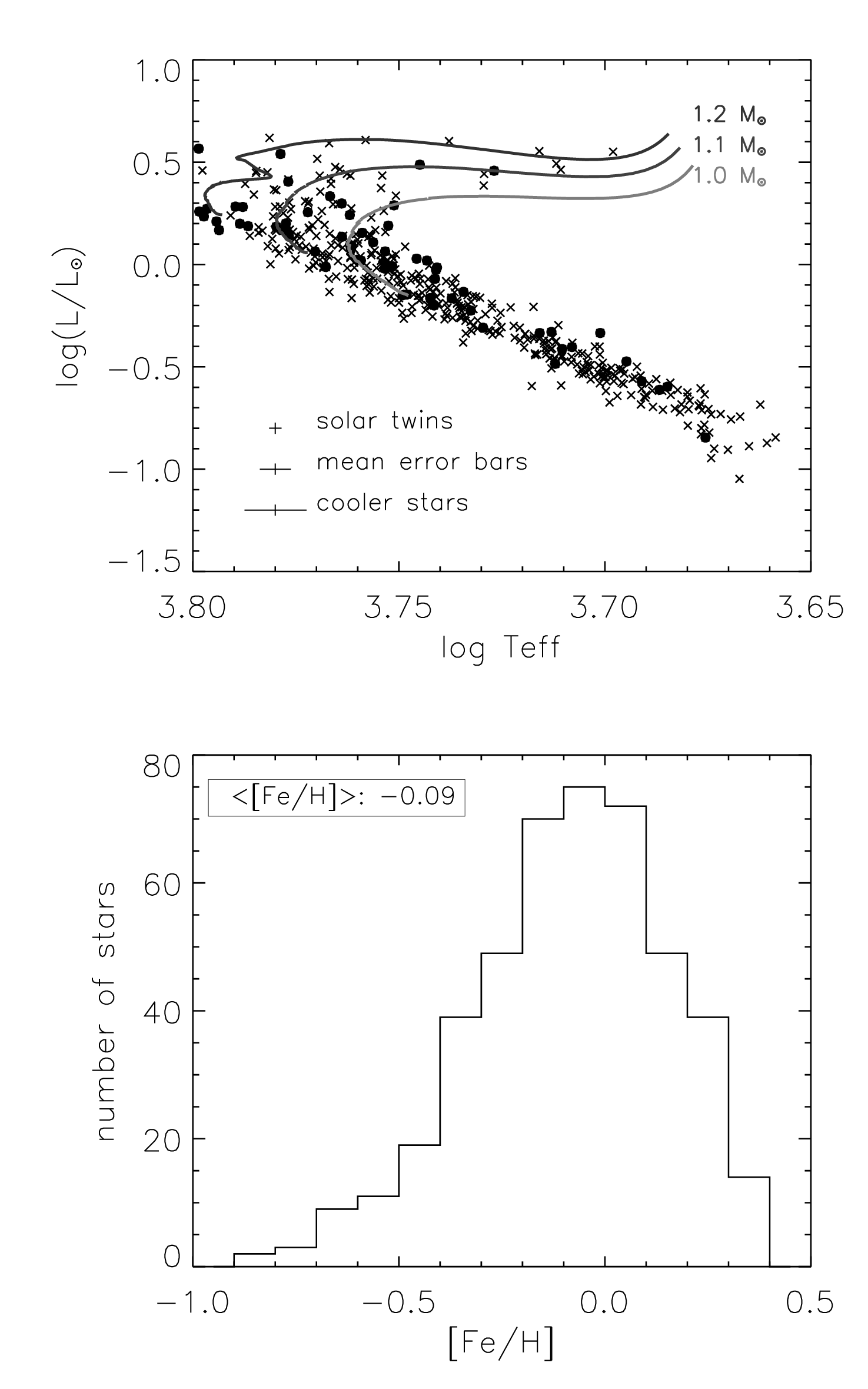}
\caption[]{In the top panel, we show the distribution of the sample stars in the H-R diagram. The filled circles represent the planet hosts in our sample. We also plot some evolutionary tracks computed with CESAM for a 1.0, 1.1 and 1.2 M$_{\sun}$. In the bottom panel, we present the metallicity distribution of the sample.}
\label{fig_analise}
\end{figure}

The luminosity was computed by considering the Hipparcos parallaxes, V magnitude and the bolometric correction. Its error was derived based on the parallax errors from Hipparcos, which was the main source of uncertainty in the calculation of luminosity. The error in the luminosity is available in the electronic table only.


\section{The spectroscopic catalogue of the HARPS GTO ``high precision'' sample}
\begin{figure*}[hth]
\centering
\includegraphics[width=17cm]{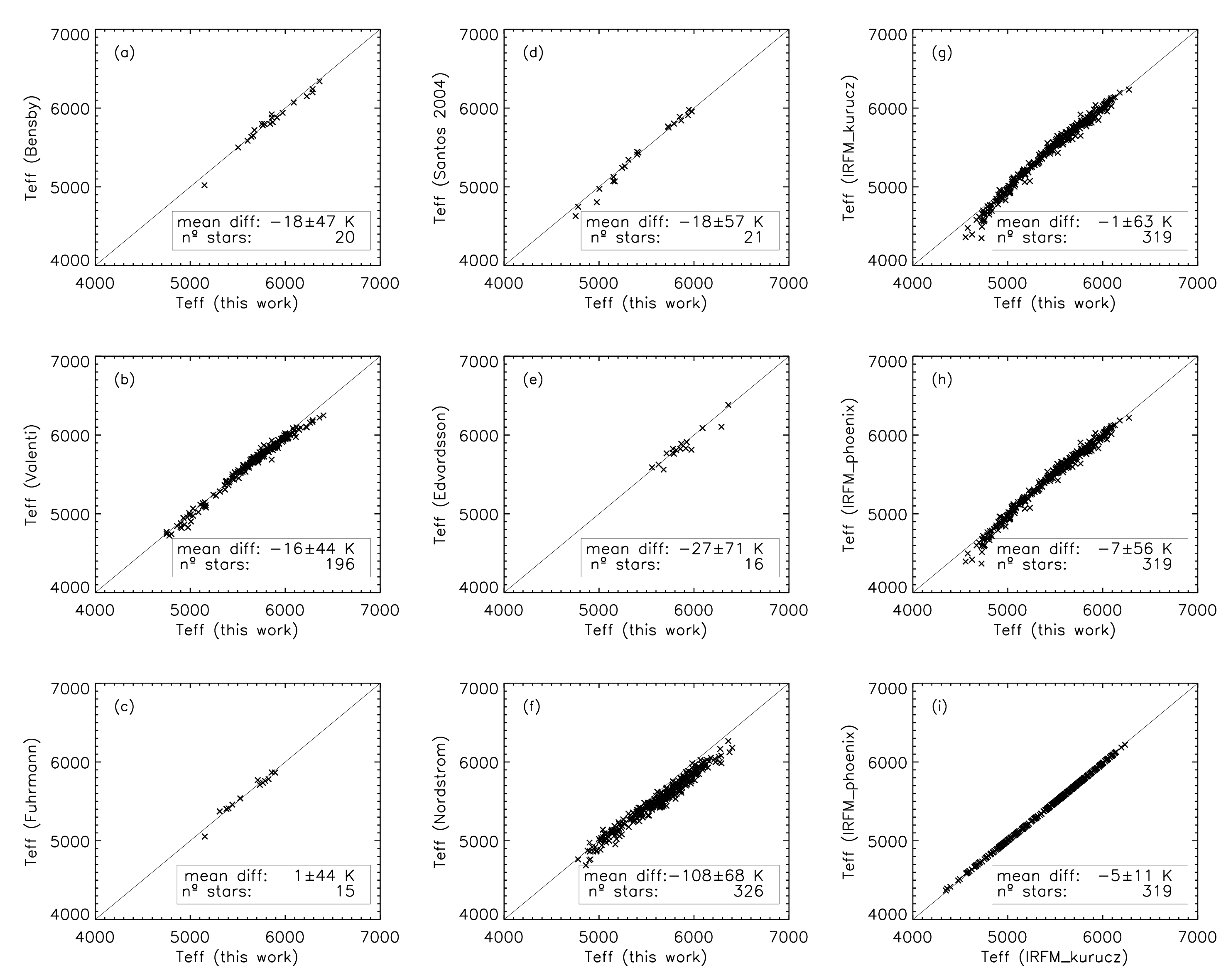}
\caption[]{Comparison of our spectroscopic results for the effective temperature with other values found in the literature. We also show the comparison with IRFM using either the Kurucz or Phoenix models.}
\label{fig_spec_teff}
\end{figure*}

Figure \ref{fig_analise} presents some characteristics of the sample. The top plot shows the distribution of the sample on the Hertzsprung-Russell diagram, where we represent evolutionary tracks for 1.0, 1.1, and 1.2 $M_{\sun}$, computed using the CESAM code \citep[][]{Morel-1997, Marques-2008} \footnote{http://www.astro.up.pt/corot/models/}. Moreover, we present the typical error boxes on this specific diagram, where the error in the luminosity is dependet mostly on the error in the parallax of each star, and the error in the temperature is derived from our spectroscopic method. This plot shows that the sample is composed mainly of main-sequence solar-type stars. In the bottom plot, we present the metallicity distribution that has a mean value of about $-0.09$ dex. This is compatible with the comparison samples presented in the work of \citet[][]{Santos-2004b}, when taken into consideration the amount of planet hosts (more metal-rich on average) that were added to the original HARPS GTO ``high precision'' sample.

Table \ref{tab3} shows a sample of the catalogue with some of the stellar parameters that have been determined. This catalog is electronically available on CDS and also at the author webpage\footnote{http://www.astro.up.pt/$\sim$sousasag/harps\_gto\_catalogue.html}.

\section{Comparison with previous works}

We compared our results with those of other authors to acess the consistency between different methods, using spectroscopy or photometry to derive stellar parameters. We used the spectroscopic results of \citet[][]{Edvardsson-1993}, \citet[][]{Bensby-2003}, \citet[][]{Valenti-2005}, \citet[][]{Santos-2004b}, \citet[][]{ Fuhrmann-2004}, and the photometric results of \citet[][]{Nordstrom-2004}. We also checked the consistency with results derived using the Infra-Red Flux Method (IRFM), following the work of \citet[][]{Casagrande-2006}.

From this point on, all differences presented are relative to our own work, i.e. ``other works''$-$``this work''.

\subsection{Effective Temperature}

For the effective temperature the results of the comparison, presented in Fig. \ref{fig_spec_teff}, are compatible with other spectroscopic studies. We found a mean difference of only $-18 \pm 47$ K for the 20 stars in common with the work of \citet[][]{Bensby-2003}. These authors used the same procedure as this paper for determining the effective temperature, but in their case for different stellar-atmosphere models (Uppsala MARCS code). In their work, they measured equivalent widths from spectra taken with the FEROS spectrograph on the 1.52\,m ESO telescope.

\citet[][]{Fuhrmann-2004} used a different procedure for the determination of the effective temperature, based on hydrogen line profiles. However, our and their results are in excellent agreement for the 15 stars that we have in common, for which we find a mean difference of $+1 \pm 44$ K.

\citet[][]{Valenti-2005} presented stellar parameters for 1040 F, G, and K dwarfs. We were able to complete a comparison using data for a larger number of stars and found good agreement with a mean difference $-16 \pm 44$ K in the effective temperature. However, a small offset can be seen in the plot for higher temperatures ($T_{\mathrm{eff}} > 6000$~K). These authors used spectral synthesis for data taken with HIRES, UCLES, and the Lick Observatory Hamilton Echelle Spectrometer.

\begin{figure*}[htp]
\centering
\includegraphics[width=17cm]{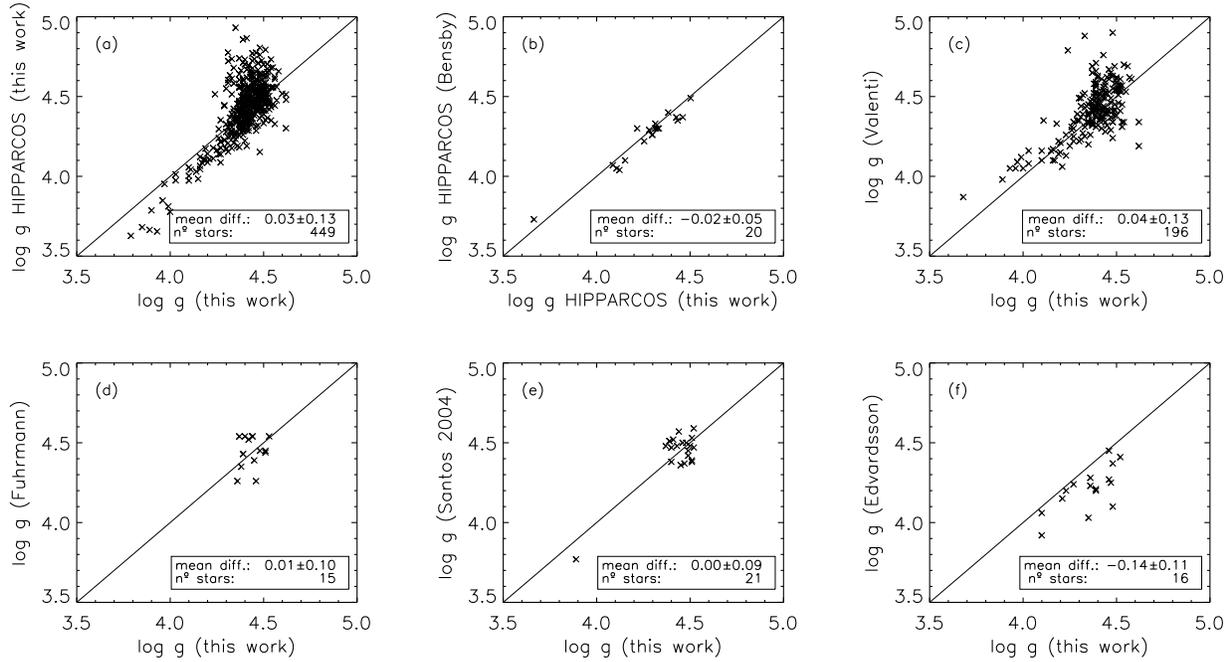}
\caption[]{Comparison of our spectroscopic results for surface gravity with other measurements in the literature.}
\label{fig_logg}
\end{figure*}

We compared our measurements derived here with our own previous results \citep[][]{Santos-2004b}, which were obtained using the same procedure but a different line list. The comparison indicated good agreement for effective temperatures with a mean difference of $-18 \pm 57$ K. For this specific comparison, most stars in common with the sample of \citet[][]{Santos-2004b} were observed with the CORALIE spectrograph and only a few with the FEROS spectrograph.

In their careful spectroscopic analysis of 189 nearby field F and G disk dwarfs, \citet[][]{Edvardsson-1993} determined effective temperatures using  photometry. For the stars in common, we found very similar effective temperatures, with a mean difference of only $-27 \pm 71$~K. 

We also compared our effective temperatures with those in \citet[][]{Nordstrom-2004} that used the IRFM--based calibration of \citet[][]{Alonso-1996}. There is a systematic deviation between our results and those of \citet[][]{Nordstrom-2004}, with a mean difference of $-108 \pm 68$~K. In particular, a clear offset appears to be present for effective temperatures above $5500$~K. We do not know the reason for this, but it is interesting to note that the \citet[][]{Alonso-1996} temperature scale is based on the Vega absolute calibration derived by \citet[][]{Alonso-1994}. In that work, it was found that the Vega absolute calibration derived from hot ($T_{\mathrm{eff}} > 6000$~K) and cool ($T_{\mathrm{eff}} < 5000$~K) stars differed by a few percent, which implied that effective temperatures for cool and hot stars were not on exactly the same scale. The Vega absolute calibration used by \citet[][]{Alonso-1996} was a weighted average of that for both hot and cool stars: it is in fact interesting to note that the offset in Fig. \ref{fig_spec_teff} appears approximately where the method showed some problems.

\subsection{Effective Temperature - The IRFM}

Casagrande et al. (2006) implemented a revised form of the IRFM, as described in their paper. They found good agreement with various spectroscopic temperature scales; they explained a $\sim100$~K systematic difference with respect to previous implementation of the IRFM, as being most likely due the adoption of the new Vega zero points and absolute calibration of 2MASS \citep[][]{Cohen-2003}. The comparison with a large set of solar analogs confirm that the temperature scale is calibrated well for solar-type stars \citep[][]{Casagrande-2006}.

The main concept behind the IRFM is to assume that the ratio between the bolometric and infrared monochromatic fluxes is a sensitive indicator of effective temperature. The bolometric flux is recovered using multi--band photometry, right across the optical and near--infrared. The flux outside the photometric filters is estimated using model atmospheres, and the infrared monochromatic fluxes are computed from 2MASS $JHK_K$ photometry. Accurate infrared photometry is crucial for recovering effective temperatures precisely and only stars with total photometric errors (i.e. ``j\_''$+$``h\_''$+$``k\_msigcom'' as given from 2MASS) smaller than 0.10~mag are used.

For 66 stars in the HARPS GTO sample we obtained accurate Johnson--Cousins $BV(RI)_C$ colors from the General Catalogue of Photometric Data \citep[][]{Mermilliod-1997} and $JHK_S$ from 2MASS, and executed the IRFM as described in \citet[][]{Casagrande-2006}. To increase the number of stars for which it was possible to apply the IRFM, we tested the results only if Johnson $BV$ and 2MASS $JHK_S$ magnitudes were used. We found the average difference to be $\Delta T_{eff} = 5 \pm 15$~K, implying that the IRFM could be applied to all HARPS GTO stars which with {\it Hipparcos} Johnson photometry \citep[][]{ESA-1997}. We also checked that if {\it Hipparcos} Tycho $B_TV_T$ magnitudes, transformed onto the Johnson system \citep[][]{Bessell-2000}, were used instead that the differences in our results would be negligible.


\begin{figure}[htp]
\centering
\includegraphics[width=6.5cm]{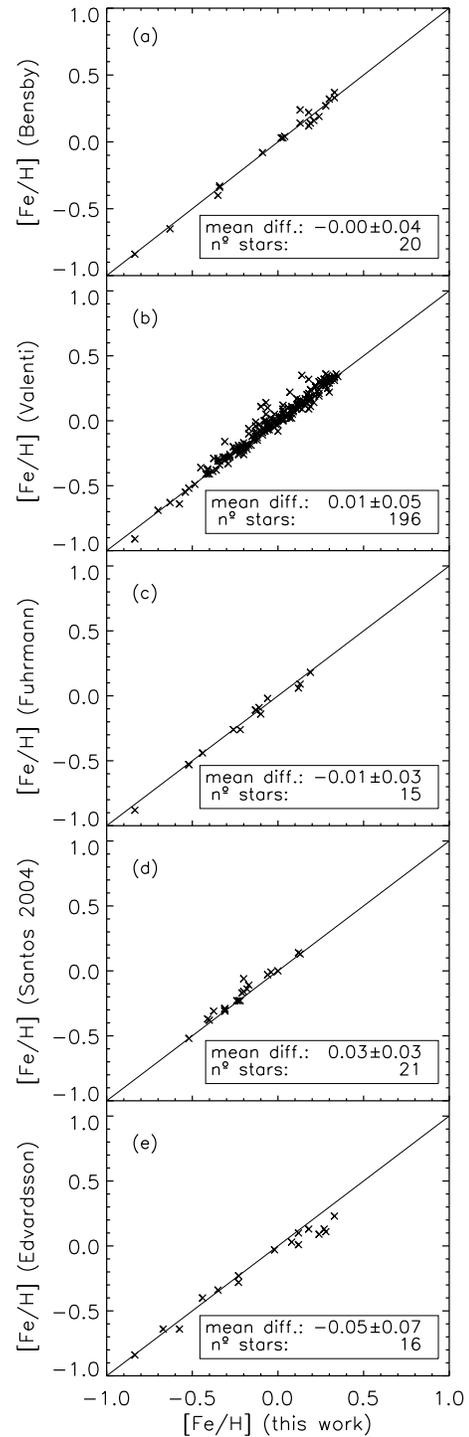}
\caption[]{Comparison of our spectroscopic measurements of [Fe/H] to others found in the literature.}
\label{fig_feh}
\end{figure}

Figure \ref{fig_spec_teff} (three right plots) shows the comparison with effective temperatures derived by the spectroscopic analysis described in Sect. 3. The agreement is excellent, corresponding to a mean difference of $-1 \pm 63$~K. 
We measure a slight disagreement when we consider cooler stars alone, which may due to inconsistencies in stellar model atmospheres at the lowest temperatures covered by the present study. In the case of the IRFM, we use the ATLAS9--ODFNEW models \citep[][]{Castelli-2003} to recover the bolometric flux that is missing from our multi--band photometry. We also tested the use of the latest Phoenix models \citep[][]{Brott-2005}, which include an extensive molecular-line list, which is particularly important for cooler stars. Using the IRFM, we found that results derived using the ATLAS9--ODFNEW models agree incredibly well with those for the Phoenix models, corresponding to a mean difference in temperature of only $-5 \pm 11$~K.
Such agreement is in part due to our implementation of the IRFM, which uses more observational data and highlights the high degree of consistency between physical descriptions in independent spectral synthesis codes.


The comparison with the IRFM in Fig. \ref{fig_spec_teff} implies that our spectroscopic method decreases its accuracy when applied to stars cooler than $\sim 4800$~K. This conclusion is expected since the method described in Sect. 3 is based on a differential analysis with respect to the Sun; it is also supported by considering the error determination for the stars: since errors are based on the distribution of the abundance determination, the dispersion is larger for cooler stars. Above $4800$~K, we find, however, very good agreement.

\subsection{Surface Gravity}

Figure \ref{fig_logg} shows the comparison between surface gravity measurements derived in our work and those from other authors. We also computed spectroscopy surface gravities using Hipparcos parallaxes and the estimated masses of stars as it is described in \citet[][]{Santos-2004b}. For the comparison between these computed gravities and the spectroscopic gravities it is observed a dispersion and offset for lower gravities. This is probably due to uncertainties in the estimated value of stellar mass, which is required in this approach to compute the gravity. The large mass uncertainty can be attributed to the considerable number of degeneracies in stellar interior models: it is possible, for example, to describe the data for a star in the H-R diagram using several models with different initial conditions \citep[see e.g.][]{Fernandes-2003}. It is interesting that some high gravity values (close to 5.0 dex) are computed using the parallaxes. These high values are not expected, since we are dealing with dwarf main sequence stars.

The surface gravities presented by \citet[][]{Bensby-2003} were determined using the stellar mass and parallax of stars. In Fig. \ref{fig_logg}, we compare the values of $\log g$ from \citet[][]{Bensby-2003} with our surface gravities computed from parallaxes. We observe perfect agreement with a mean difference of $-0.02 \pm 0.05$ dex.

We also find an excellent agreement with the surface gravities determined by \citet[][]{Fuhrmann-2004} in addition to our old results using the same method but a smaller line list \citep[][]{Santos-2004b}.

\citet[][]{Edvardsson-1993} measured the surface gravity using a different method, which used the Balmer discontinuity index. When comparing their values with our spectroscopic surface gravities, we find a rather large deviation with a mean difference of $-0.14 \pm 0.11$ dex.

The mean difference between values obtained by \citet[][]{Valenti-2005}, using spectral synthesis, and ourselves is only $0.04 \pm 0.13$ dex, which indicates good agreement between the measurements. There is, however, significant dispersion in this comparison. For a few stars, \citet[][]{Valenti-2005} derive unusually high values of gravity; in the lower regime of gravity, they also measure higher values than the ones we present in this work. A comparison between \citet[][]{Valenti-2005} gravities and independent computation of gravity using Hipparcos parallaxes reveals a larger offset for smaller gravities. We note that this does not imply that gravities given by these authors are incorrect, but that the offset is greater than for our values when comparion to the computed gravities using parallaxes. We also recall that the mass estimates have a large uncertainty, which may contribute to the observed offsets.

\subsection{Metallicity}

Figure \ref{fig_feh} shows the comparison of the metallicity estimates in this work with those derived in other works.
As for the other parameters, all metallicities agree well within the errors. The results from \citet[][]{Edvardsson-1993} differ the most from our values (with only 0.05 dex), in particular for metallicities above $\sim$$0.1$\,dex.


\section{A calibration of the effective temperature as a function of \textit{B-V} and [Fe/H]}

\begin{figure}[htp]
\centering
\includegraphics[width=8cm]{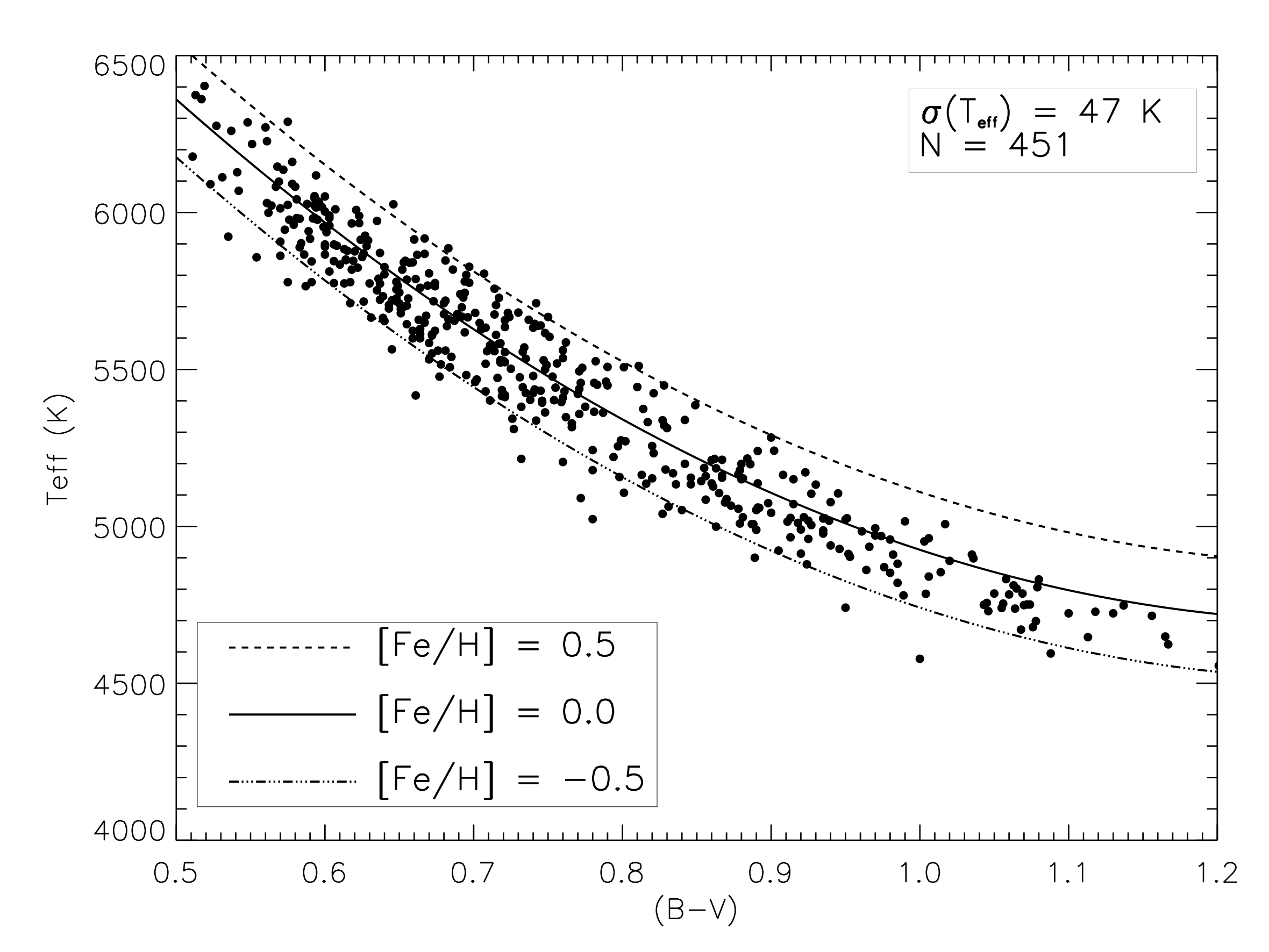}
\caption[]{Calibration of the effective temperature as a function of the color index \textit{B-V} and [Fe/H]. The 3 fitted lines correspond to lines with constant values of [Fe/H] (-0.5, 0.0, 0.5). }
\label{fig_teff_bv_feh}
\end{figure}

Using the parameters presented in Table \ref{tab3} and the $B-V$ value taken from the Hipparcos catalogue \citep[][]{ESA-1997}, we derived a new calibration of the effective temperature as a function of \textit{B-V} and [Fe/H]. The result is illustrated in Fig. \ref{fig_teff_bv_feh} and the calibration is expressed by:


\begin{equation}
 T_{\mathrm{eff}} = 9114 - 6827(B{-}V) +2638(B{-}V)^2 + \text{368[Fe/H].}
\end{equation}

The standard deviation of the fit is only 47 K, illustrating the quality of the relation. Such a result was already observed in a calibration completed by \citet[][]{Santos-2004b}, using a much smaller number of stars. This calibration can be useful and applied to stars without the need for a detailed spectroscopic analysis, with the guarantee that the result will be in the same effective temperature scale.
This calibration is valid in the following intervals: $4500 K < T_{eff} <6400 K$, $-0.85 < [Fe/H] < 0.40$, and $ 0.51 < B{-}V < 1.20$. The small dispersion also attest to the quality of the metallicity values derived in this work.

\section{Planet-hosts in the sample}



\begin{figure}[htp]
\centering
\includegraphics[width=8.5cm]{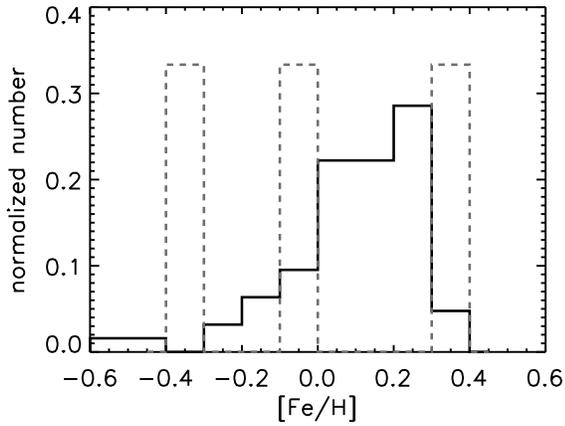}
\caption[]{Metallicity distribution of jovian planet hosts and neptunian planet hosts within the HARPS GTO ``high precision'' spectroscopic catalogue.}
\label{fig_planets2}
\end{figure}

The well-established correlation between the presence of a giant planet and the metallicity of its host star \citep[][]{Santos-2004b,Fischer_Valenti-2005} is illustrated by our results. This trend is observed in Fig. \ref{fig_planets2}, where the metallicity
distribution of jovian-like planets is clearly located towards the metal-rich regime. The subsample of 66 planet hosts has 
a mean metallicity of $+0.09$ dex. For comparison, the sample presented in this work has a mean 
value of $-0.12$ dex, if we only include stars without planets, and $-$0.09, if we include all the stars (see also Fig.\,\ref{fig_analise}). 

From the 66 planet hosts belonging to our sample, three host neptunian planets (HD4308, HD69830 and HD160691). The first two of these only harbor
neptunian planets \citep[][]{Udry-2006,Lovis-2006}, while the third also has Jupiter-like planets \citep[][]{Pepe-2007}. Others Neptune-like planets were discovered with HARPS, but do not belong to this sample. In Fig.\ref{fig_planets2}, we present the metallicity distribution of these two types of host stars. Althought the numbers are small, we find a wider spread of metallicities for stars hosting Neptunian-like planets.

In Table \ref{tab4}, we compile a list of [Fe/H] values for all known planet hosts orbited by neptunes or super-earths. This list was compiled using the extrasolar-planets encyclopaedia (http://www.exoplanet.eu), for which we select the discovered planets with masses ($M_p\sin i$) less than 25\,M$_\oplus$. In the table, we indicate whether each star also hosts jovian planet(s) (``yes''), or if it only hosts neptunian planets or super-earths (``no''). The [Fe/H] values are listed in the table for each star. 


To increase the number of neptunian-planet hosts, we must include M stars presented in the table. We note that the metallicity determination of M stars can be difficult; we therefore use the mean of the [Fe/H] values quoted in the literature.

We should mention that we do not expect the inclusion of M-dwarfs in our comparison to strongly compromise or bias our conclusions. First, the metallicity distribution of M-dwarfs in the solar neighborhood differs only slightly from that of field FGK-dwarfs. The offset on the metal distribution is small ($\sim$0.07\,dex), and possibly within the margins of errors \citep[][]{Bonfils-2005}. Secondly, the fact that these metallicities are more uncertain will produce a scatter, but likely not a systematic offset, in the [Fe/H] distribution of stars with Neptune-like planets. Finally, it is reasonable to assume that the influence of metallicity on planet-formation efficiency is independent of either host mass or spectral type.



\begin{figure}[htp]
\centering
\includegraphics[width=8.5cm]{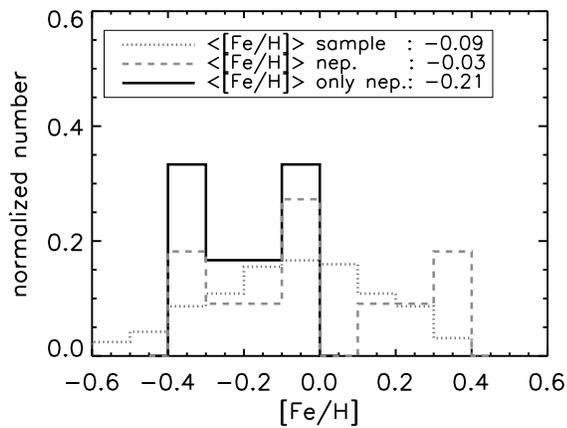}
\caption[]{Metallicity distribution for the sample presented in this work (dotted line), for stars hosting neptunian planets (dashed line), and for stars exclusively hosting neptunian planets (full line). The mean metallicity of each distribution is indicated on the plot.}
\label{fig_planets3}
\end{figure}

Using the values of [Fe/H] listed in Table \ref{tab4}, we compared the metallicity distribution of the spectroscopic catalogue presented in this paper, which corresponds to the [Fe/H] distribution of the solar neighborhood, with two different distributions. The first, composed of all stars presented in Table\,\ref{tab4}, and the second, composed only of stars that do not host jovian planets (``no'' in the table). The result can be observed in Fig.\,\ref{fig_planets3}. The metallicity distribution of the catalogue presented in this work has a mean metallicity of $<[Fe/H]>=-0.09$, while the metallicity distribution of stars known to harbor at least one neptunian planet corresponds to a mean metallicity of $<[Fe/H]>=-0.03$. Finally, stars known to host only neptunian planets have a mean metallicity of $<[Fe/H]>=-0.21$.

We performed a Kolmogorov-Smirnov  test to check the probability that samples are part of the same population. We compare the sample composed of all stars in Table\,\ref{tab4}, to that composed only of giant-planet hosts in the catalogue. According to a Kolmogorov-Smirnov test, there is a probability of ~3\% that the two samples belong to the same population. We applied the same test to stars that host only neptunian planets (``no'' in the table) and found a probability of only 0.5\%. This result, although preliminary, implies that these samples are unlikely to belong to the same population.

Although the sizes of the datasets are small, we attempt to estimate a lower limit value of the frequency of Neptune-like planets (or super-earths) found, as a function of stellar metallicity. For each of the three metallicity bins $[Fe/H]<-0.15$, $-0.15<[Fe/H]\le0.15$, and $[Fe/H]\ge0.15$, we computed the ratio of the number of stars with Neptune-like planets in Table \ref{tab4} to the number of stars in our entire sample. In this estimate, we assume that the metallicity distribution of the 451 stars represents the metallicity distribution of the planet search samples used to find the planets listed in Table \ref{tab4}.

The results show that the frequency of stars hosting neptunians is 1.8\%, 2.0\%, and 5.1\%, respectively for each of the metallicity bins mentioned above. These numbers may, however, be biased towards the high-metallicity regime, since several stars in Table \ref{tab4} also have jovian planets. 
We further note that we attempted to find Neptune-like planets orbiting some of the FGK stars in Table\,\ref{tab4} because of their high metallicity \citep[e.g.][]{Melo-2007}. In other cases, the approach taken to detect a Neptune-like planet assumed that the star hosted already at least one jupiter-like companion \citep[e.g.][]{Santos-2004a,McArthur-2004}. If we consider stars hosting $only$ Neptune-like planets, these frequencies change to 1.8, 1.5, and 0.0\% (there are no stars that host Neptune-like planets, that do not host Jupiter-like planets with [Fe/H]$>-$0.06).


\begin{table}[t]
\caption[]{Neptunian hosts and their metal content. We also list the mass (in earth masses) of the least massive planet 
orbiting the star, as well as an indication of whether there is one or more jovian companions.}
\begin{tabular}{lrccl}
\hline
\hline
\noalign{\smallskip}
star & [Fe/H] & jov.? & $M_p\sin i$ & [Fe/H] reference\\
\hline
\object{Gl 581 }   & -0.33 &  no &   5.09 & \citet[][]{Bean-2006}\\
                   & -0.25  &     &        & \citet[][]{Bonfils-2005b}\\
                   & -0.29  &     &        & Mean value\\
\object{Gl 876 }   & -0.12 & yes &   5.72 & \citet[][]{Bean-2006}\\
                   & -0.03  &     &        & \citet[][]{Bonfils-2005b}\\
                   & -0.08  &     &        & Mean value\\
\object{HD69830}   & -0.06 &  no & 10.49  & This work\\
\object{HD160691}  &  0.30 & yes & 13.99  & This work\\
\object{Gl 674 }   & -0.28 &  no & 11.76  & \citet[][]{Bonfils-2007}\\
\object{55 Cnc e}  &  0.33 & yes & 10.81  & \citet[][]{Santos-2004b}\\
\object{HD4308}    & -0.34 &  no & 14.94  & This work\\
\object{HD190360}  &  0.24 & yes & 18.12  & \citet[][]{Santos-2004b}\\
\object{HD219828}  &  0.19 & yes & 20.98  & \citet[][]{Melo-2007}\\
\object{GJ 436}    & -0.32 &  no & 22.89  & \citet[][]{Bean-2006}\\
                   &  0.02 &     &        & \citet[][]{Bonfils-2005b}\\
                   & -0.15 &     &        & Mean value\\
\object{Gl 176}    & -0.10 &  no & 24.16  & \citet[][]{Endl-2007}\\

\hline
\end{tabular}
\label{tab4}
\end{table}



These numbers imply that the few low-mass planets found appear to follow a trend in metallicity that differ from well-established relations for giant planets first pointed out by \citet[][]{Udry-2006}, this result, is supported by planet-formation models based on the core-accretion paradigm \citep[][]{IdaLin-2004, Benz-2006}. These works suggest that neptunian planets should be found in a wider range of stellar metallicities. Lower-mass planets could even be preferentially found orbiting metal-poorer stars. 


Using the numbers presented in Table\,\ref{tab4}, we measure a preliminary value for the 
Jupiter-to-Neptune ratio as a function of stellar metallicity. We estimate the value using the metallicity distribution
of stars with only jovian planets in our HARPS sample, and compared it with the same distribution
for stars orbited only by Neptune-like planets. For this, we used the sample of three stars 
mentioned above (HD4308, HD69830 and HD160691, belonging to the sample presented in this paper -- case\,1), 
all stars $only$ with neptunian planets listed in Table\,\ref{tab4} (case\,2), and all stars in the table (case\,3). 

The results show that in the same three metallicity intervals mentioned above, the Jupiter-to-Neptune ratios 
are, by order of increasing metallicity, 5:1, 28:1, and 30:1 (case\,1), 5:3, 28:3, 30:0 (case\,2), and 5:3 28:4 30:4 (case\,3). 
In other words, this result tentatively suggests that the Jupiter-to-Neptune ratio may be higher at higher metallicities.
This is expected by models of planet formation based on the core-accretion process \citep[][]{Benz-2006}.

These final results should be interpreted with caution, however, since most of the stars that have $only$ with Neptune-like planets listed
in Table\,\ref{tab4} are M-dwarfs. It has been proposed that M-dwarf stars may have a lower rate of Jupiter-like planets, but (at least in the short planet period range) a high incidence of Neptune-like planet detections \citep[][]{Endl-2006,Bonfils-2007}.
The above mentioned bias in the case of a search for neptunian planets around metal-rich (or planet-host) stars may also
induce important biases in this analysis.

\section{Summary}

We have measured accurate stellar parameters for a well-defined sample of solar-type stars, which were originally observed as a part of a search for very low-mass planets. We have presented a catalog of stellar parameters for 451 stars which is available online. 
We compared our results with other works in the literature. The effective temperatures obtained were also compared 
with values derived using IRFM. All comparisons indicated a very good agreement in despite of the different 
methods used by the different authors.

A new calibration was developed for the determination of effective temperature using the index color $B-V$ and metallicity. 

The results presented here were explored to study the metallicity-planet correlation, in particular
for stars hosting planets in the Neptune-mass range. We showed that, when compared with their higher-mass counterparts, 
neptunian planets did not follow the same trend. Low-mass planets appeared to be formed in a lower metal content 
environment. Interestingly, the Jupiter-to-Neptune ratio was found to be an increasing function of stellar metallicity.
The results presented in this paper have shown that the study of the chemical abundances of Neptune-host stars
provides new important constraints on models of planet formation and evolution.

\begin{acknowledgements}
S.G.S and N.C.S. would like to acknowledge the support from the Funda\c{c}\~ao para a Ci\^encia e Tecnologia (Portugal) in the form of fellowships and grants SFRH/BD/17952/2004, POCI/CTE-AST/56453/2004, PPCDT/CTE-AST/56453/2004 and PTDC/CTE-AST/66181/2006, with funds from the European program FEDER. We thank the anonymous referee for the useful comments and suggestions.
\end{acknowledgements}

\bibliographystyle{aa}
\bibliography{sousa_bibliography}

\end{document}